\documentclass[aps,prl,twocolumn,superscriptaddress]{revtex4}

\usepackage{epsf,epsfig,amsmath,amssymb,amsfonts}
\usepackage{color}
\usepackage{slashed}

\newsavebox{\ns}
\newsavebox{\dbrane}

\def\be{\begin{equation}}
\def\ee{\end{equation}}
\def\bea{\begin{eqnarray}}
\def\eea{\end{eqnarray}}

\def\Dslash{\,\,{\raise.15ex\hbox{/}\mkern-12mu D}}
\def\Dbarslash{\,\,{\raise.15ex\hbox{/}\mkern-12mu {\bar D}}}
\def\delslash{\,\,{\raise.15ex\hbox{/}\mkern-9mu \partial}}
\def\delbarslash{\,\,{\raise.15ex\hbox{/}\mkern-9mu {\bar\partial}}}
\def\pslash{\,\,{\raise.15ex\hbox{/}\mkern-9mu p}}
\def\calDslash{\,\,{\raise.15ex\hbox{/}\mkern-12mu {\cal D}}}

\newcommand\diff{\mbox{d}}

\newcommand{\nn}{\nonumber \\}

\newcommand{\dd}{\diff}

\allowdisplaybreaks

\begin{document}

\title{Warped AdS$_3$, dS$_3$ and flows from $\mathcal{N} = (0,2)$ SCFTs}

 \author{Eoin \'O Colg\'ain}
 \affiliation{C.N.Yang Institute for Theoretical Physics, SUNY Stony Brook, NY 11794-3840, USA}
 \affiliation{Department of Mathematics, University of Surrey, Guildford GU2 7XH, UK}

\preprint{YITP}

\begin{abstract}
We present the general form of all timelike supersymmetric solutions to 3D U(1)$^3$ gauged supergravity, a known consistent truncation of string theory. We uncover a rich vacuum structure, including an infinite class of  new timelike-warped AdS$_3$ (G\"{o}del) and timelike-warped dS$_3$ critical points.  We outline the construction of supersymmetric flows, driven by irrelevant scalar operators in the SCFT, which interpolate between critical points. For flows from AdS$_3$ to G\"{o}del, the natural candidate for the central charge decreases along the flow. Flows to timelike-warped dS$_3$ exhibit topology change.  \end{abstract}

\maketitle

\setcounter{equation}{0}

\section{Introduction} \label{Introduction} 
It is a remarkable fact that the geometrical Bekenstein-Hawking (BH) entropy of black holes with AdS$_3$ near-horizons can be derived from the central charge of a two-dimensional (2D) CFT \cite{Strominger:1996sh, Strominger:1997eq}. This result, a key precusor to AdS/CFT \cite{Maldacena:1997re},  rests on the Brown-Henneaux analysis of the asymptotic symmetries of AdS$_3$ \cite{Brown:1986nw} and the Cardy formula \cite{Cardy:1986ie}, which permits one to determine the asymptotic density of states in a CFT in the semi-classical limit. 

It is well-known that Kerr black holes, candidates for astrophysical black holes, e.g. Cygnus X-1 \cite{Gou:2013dna}, exhibit \textit{warped} AdS$_3$ near-horizons \cite{Bengtsson:2005zj}. In recent years, the matching of BH entropy through the Cardy formula lead to a bold conjecture that there is a (warped) CFT dual to Kerr black holes \cite{Guica:2008mu} (see \cite{Compere:2012jk} for a review). A greater understanding of the putative dual QFT, if it is even a CFT \cite{ElShowk:2011cm,Song:2011sr, Compere:2013bya,Detournay:2012pc, Compere:2014bia}, requires a theory with a UV completion, such as string theory.   

In this work we take a step in this direction by identifying warped AdS$_3$ vacua of $\mathcal{N}=2$ U(1)$^3$ gauged supergravity \cite{Karndumri:2013iqa}, a consistent truncation of string theory \cite{Cvetic:1999xp, Colgain:2014pha}, and offering evidence that they can be connected  to understood AdS$_3$ vacua by supersymmetric flows. This places holography on a firmer footing, since at one end of the flow, the supersymmetric AdS$_3$ vacua are dual to 2D $\mathcal{N} = (0,2)$ SCFTs \cite{Maldacena:2000mw,Almuhairi:2011ws,Benini:2012cz,Benini:2013cda}, whose central charge and R symmetry can be determined exactly using  $c$-extremization \cite{Benini:2012cz,Benini:2013cda} and agree with holographic calculations (see \cite{Baggio:2014hua} for subleading terms).  

Following a review in the next section, we make the following novel contributions. Firstly, we present the general form - dictated by the bps conditions - of all supersymmetric timelike solutions to 3D U(1)$^3$ gauged supergravity, including an infinite class of new half-bps critical points, going under the moniker timelike-warped AdS$_3$ (G\"{o}del) and timelike-warped dS$_3$ in the literature. Indeed, the latter is a known solution to Topologically Massive Gravity with a positive cosmological constant \cite{Anninos:2009jt} and here we provide potentially the first example in both a supersymmetric and string theory context. Being timelike-warped, the geometries exhibit characteristic closed-timelike-curves (CTCs), signaling a breakdown in unitarity in the dual theory.  Along with the G\"{o}del universe \cite{Godel}, which is not ruled out by supersymmetry \cite{Gauntlett:2002nw, Harmark:2003ud}, a version of Hawking's Chronology Protection Conjecture \cite{Hawking:1991nk} is expected for timelike-warped dS$_3$. See \cite{Herdeiro:2000ap,Drukker:2003sc,Caldarelli:2004mz, Raeymaekers:2011uz} for related works in the AdS/CFT context. 

We construct numerical supersymmetric flows from AdS$_3$ to timelike-warped critical points and identify the flows as deformations of the 2D SCFT by irrelevant scalar operators. We show that the inverse of the real superpotential monotonically decreases along flows to timelike-warped AdS$_3$ vacua and calculate an expression for the candidate central charge in terms of twist parameters. For flows to timelike-warped dS$_3$, the curvature of the Riemann surface changes sign and the topology changes. Since the 2D SCFTs correspond to twisted compactifications of $\mathcal{N}=4$ super-Yang-Mills, our 3D flows can be uplifted to 5D, where they may be interpreted as deformations of $\mathcal{N} =4$ super-Yang-Mills. This short letter highlights the existence of novel warped critical points; further examples of supersymmetric flows, the 5D uplift and the generalisation to null spacetimes can be found in \cite{Colgain:2015mta}. 

\section{3D U(1)$^3$ gauged supergravity}
We consider 3D $\mathcal{N}=2$ gauged supergravity \cite{Karndumri:2013iqa}, which uplifts on a constant curvature Riemann surface of genus $\frak{g}$, $\Sigma_{\frak{g}}$, to well-known 5D U(1)$^3$ gauged supergravity, where it can be further embedded consistently in higher dimensions \cite{Cvetic:1999xp, Colgain:2014pha}.  Examples of consistent truncations of string theory with warped AdS$_3$ vacua have appeared previously in \cite{Detournay:2012dz} (see also \cite{Karndumri:2013dca}). 

The action for the theory may be written as  
\bea
\label{Einsteinact}
\mathcal{L}_3 &=&  R *_3 \mathbf{1} - \frac{1}{2} \sum_{i=1}^3 \left[ \dd W_i \wedge *_3 \dd W_i + e^{2 W_i} G^i
\wedge *_3 G^i \right]   \nn &+&
8 \left( T^2 - \sum_{i=1}^3 (\partial_{W_i} T)^2 \right) *_3 \mathbf{1} \nn
 &-& a_1 B^2 \wedge \dd B^3 - a_2 B^3 \wedge \dd B^1
- a_3 B^1 \wedge \dd B^2,  
\eea
with the field content comprising three scalars, $W_i$,  and three gauge fields, $G^i = \dd B^i$, which may be rewritten in the canonical form of 3D gauged supergravity \cite{deWit:2003ja}. $T$ denotes the superpotential 
\be
T = \sum_{i=1}^3 \left( \frac{1}{2} e^{-W_i} - \frac{a_i}{4} e^{W_i +K} \right), 
\ee
$K = - \sum_{i} W_i$ is the K\"{a}hler potential of the scalar manifold, and $a_i$, $i=1, 2, 3$ denote constants that are constrained by the curvature $\kappa$ of $\Sigma_{\frak{g}}$
\be
\label{constraint}
a_1 + a_2 + a_3 = - \kappa. 
\ee
We note there is the freedom to change the sign of $T$ and the potential does not change \footnote{This may be undertaken by changing the sign of $a_i$, with equation (\ref{constraint}) unchanged, and analytically continuing $W_i = \tilde{W}_i + i \pi$, with $\tilde{W}_i \in \mathbb{R}$. Equation (3) is unaffected.}. 

Through AdS/CFT  \cite{Maldacena:1997re}, AdS$_3$ vacua of the above supergravity correspond to 2D SCFTs arising through twisted compactifications of 4D $\mathcal{N}=4$ super Yang-Mills with gauge group U(N) on $\Sigma_{\frak{g}}$ \cite{Vafa:1994tf, Bershadsky:1995vm}. To preserve supersymmetry one ``twists" the theory by turning on gauge fields coupled to the SO(6) R symmetry of the 4D theory. For twists involving the SO(2)$^3$ Cartan subgroup of the R symmetry, the twist parameters, $a_i$, must satisfy (\ref{constraint}), a necessary condition for $\mathcal{N} =2$ supergravity.  Supersymmetry is enhanced to $\mathcal{N} = (2,2)$ and $\mathcal{N} = (4,4)$, when one or two of the $a_i$ vanish, respectively. 

Supersymmetric AdS$_3$ vacua of the action (\ref{Einsteinact}) correspond to the critical points,  $\partial_{W_i} T =0$ \cite{Karndumri:2013dca}, 
\be
\label{ads_crit}
e^{W_i} = -\frac{\prod_{j \neq i} a_j }{\kappa + 2  a_i}. 
\ee
These vacua have featured in a series of works \cite{Maldacena:2000mw,Cucu:2003bm,Almuhairi:2011ws,Benini:2012cz, Benini:2013cda}. From extrema of $T$, one can see there is no good AdS$_3$ vacuum dual to $\mathcal{N} = (4,4)$ SCFTs and that $\mathcal{N} = (2,2)$ vacua only exist when $\frak{g} > 1$.

\section{All timelike solutions} 
Given a supergravity theory, it is  feasible to invoke Killing spinor techniques to find all supersymmetric solutions, e. g. \cite{Gauntlett:2002nw, Gauntlett:2003fk, Gutowski:2004yv} in 5D.  Here we present the timelike solutions to the theory (\ref{Einsteinact}). Full details of the classification exercise appear elsewhere \cite{Colgain:2015mta}. 

The class of supersymmetric geometries is characterised by a real  timelike Killing vector $P_0$, $\mathcal{L}_{P_0} W_i = \mathcal{L}_{P_0} G^i =0$ and an additional  complex vector, $P_z =  P_1 + i P_2$. Suitably normalised, we have $P_{a} \cdot P_b  = \eta_{ab}$, $\eta_{ab} = (-1,1,1)$ and $a= 0, 1, 2$. The existence of a Killing spinor is equivalent to the differential conditions \cite{Colgain:2015mta}: 
\bea
\dd P_0 &=& 4 \, T *_3 P_0, \\
\label{diff_cond2} e^{- \frac{1}{2} K} \dd [e^{\frac{1}{2} K} P_z ] &=&  \sum_{i=1}^3 \left( e^{-W_i} *_3 + i B^i \right) \wedge P_z.
\eea
The field strengths $G^i$ are completely determined and break supersymmetry by one-half when non-zero,
\be
G^i = e^{-W_i} \left( -4 \partial_{W_i} T \, *_3 P_0 + P_0 \wedge \dd W_i \right), 
\ee
meaning $G^i =0$ at AdS$_3$ vacua. This appears to contradict the existence of well known holographic flows from AdS$_5$ to AdS$_3$ \cite{Maldacena:2000mw}, but in our conventions these fall into the null class of spacetimes, $P_a \cdot P_b = 0$. 

Given these conditions, it is a straightforward exercise  to introduce coordinates $P_0 \equiv \partial_{\tau}$, $P_z = e^{D-\frac{1}{2} K} (\dd x_1 + i \dd x_2)$, so that the spacetimes take the form: 
\bea
\dd s_3^2 &=& - (\dd \tau + \rho)^2 + e^{2 D-K} (\dd x_1^2 + \dd x_2^2), \nn
G^i &=& e^{-W_i} \biggl[ -4 \partial_{W_i} T e^{2D -K} \dd x_1 \wedge \dd x_2 \nn && \phantom{xxxxxxxx} + (\dd \tau + \rho) \wedge \dd W_i \biggr], 
\eea
where $\rho$ is a one-form connection on the Riemann surface parametrised by ($x_1, x_2$), with $\dd \rho = 4 T e^{2 D-K} \dd x_1 \wedge \dd x_2$. Here $D(x_1, x_2)$ is modulo a convenient factor of the K\"ahler potential, a warp factor parametrising the vector $P_z$, and in turn the Riemann surface. Inserting the expressions for $G^i$ into the flux equations of motion (EOMs), one can derive the following equation for the scalars: 
\bea
\label{scalarEOM}
\nabla^2 e^{W_i} = 2 e^{2D} \biggl[\frac{4 T}{e^{K}} -  \sum_{j \neq i} a_j (e^{W_j}+e^{W_i}) +  \prod_{j \neq i} a_j  \biggr]. 
\eea
From (\ref{diff_cond2}), one derives a differential condition for the warp factor, 
\be
\label{Liouville}
\nabla^2 D = 4 \sum_{i=1}^3 e^{-W_i} (\partial_{W_i} T + T ) e^{2D -K}. 
\ee
We note for a given constant value of $W_i$, this equation reduces to the Liouville equation on the Riemann surface, $\nabla^2 D = -\mathcal{K} e^{2D}$, with Gaussian curvature $\mathcal{K}$. 

Using these four equations, it is possible to show that the scalar and Einstein EOMs are satisfied. It can be independently checked that  the EOMs are consistent with the integrability conditions  \cite{Colgain:2015mta}, as expected. 

\begin{figure}[h]
\label{range}
\centering
\includegraphics[width=0.4\textwidth]{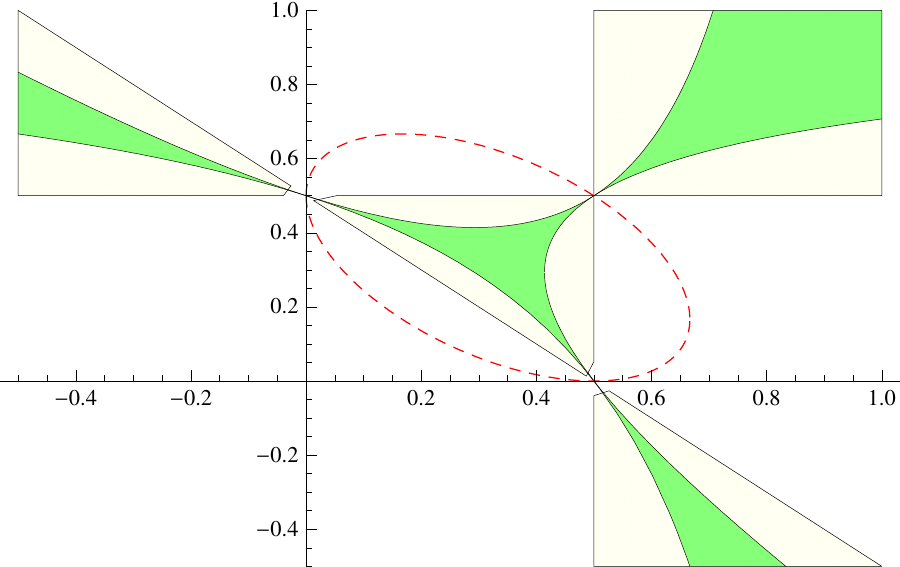}
\caption{The range of parameters in the $(a_2, a_3)$ plane where the scalars $W_i$ remain real for AdS$_3$ vacua (cream) and warped AdS$_3$ vacua (green). The dotted red line separates external ($\mathcal{K} < 0$) from internal regions ($\mathcal{K} > 0$).}
\end{figure}
\vspace{-5mm}

\section{New critical points} 
In this section, we get oriented by recovering the AdS$_3$ vacua (\ref{ads_crit}). For simplicity, we introduce a radial direction $r = \sqrt{x_1^2 + x_2^2}$ and a U(1) isometry parametrised by $\varphi$. A general solution to (\ref{Liouville}) exists where 
\be
e^{D} = \frac{2 \sqrt{|\mathcal{K}|} }{|\mathcal{K}| + \mathcal{K} r^2 }, 
\ee
resulting in a spacetime metric of the form: 
\bea
\label{timelike_metric}
\dd s^2_3 &=& - \ell^2 \left( \dd \tau - \frac{\textrm{sgn}(\mathcal{K}) r^2}{[1 + \textrm{sgn}(\mathcal{K}) r^2]} \dd \varphi \right)^2 \nn 
&& \phantom{xxxxxxxx} + \frac{e^{-K}}{|\mathcal{K}|} \left[ \frac{4 ( \dd r^2 + r^2 \dd \varphi^2)}{(1+ \textrm{sgn}(\mathcal{K}) r^2 )^2} \right].  
\eea
For AdS$_3$ vacua, $\mathcal{K} =- 4 T e^{-K} \sum_i e^{-W_i}|_{\partial_{W_i} T =0}$, which upon redefinition, $r = \tanh \rho$ and a shift $\varphi \rightarrow \varphi - \tau$, leads to the usual form of global AdS$_3$ (radius $\ell = \frac{2}{T}|_{\partial_{W_i} T =0}$), 
\be
\dd s^2_3 = \ell^2 [ -\cosh^2 \rho \dd \tau^2 + \dd \rho^2 + \sinh^2 \rho \dd \varphi^2 ].  
\ee

We now present a key observation of this letter, namely that (\ref{scalarEOM}) has a second critical point, i. e. solutions with $\partial_{a} W_i =0$, supported by fluxes. In addition to (\ref{ads_crit}), the RHS of (\ref{scalarEOM}) vanishes when 
\be
\label{new_crit}
e^{W_i}  = \sum_{j \neq i} a_j+ \frac{\kappa}{2}  + \frac{\prod_{j \neq i} a_j}{\kappa} . 
\ee
This exhausts the possibility for additional critical points beyond the supersymmetric AdS$_3$ vacuum. 
For $W_i \in \mathbb{R}$, a requirement for real solutions, necessarily $\kappa < 0$, so without loss of generality we set $\kappa = -1$. Furthermore, the range in parameter space where good vacua is constrained, as depicted in FIG. 1. From FIG. 1, suppressing $a_1$ through the supersymmetry condition $a_1 = 1-a_2-a_3$, we recognise that within the range of parameters where good AdS$_3$ vacua exist (cream), there are regions where additional new critical points exist (green). Points in parameter space where supersymmetry is enhanced to $\mathcal{N} = (2,2)$ ($\mathcal{K} =0$) e. g. $(a_2, a_3) = (\frac{1}{2}, 0), (0, \frac{1}{2}), (\frac{1}{2}, \frac{1}{2})$ on the dashed red locus are excluded, meaning new critical points only exist for $\mathcal{N} = (0,2)$ supersymmetry. 

As one crosses the dashed red locus in FIG. 1, the topology of the Riemann surface parmetrised by $(x_1, x_2)$ changes from H$^2$ externally to S$^2$ internally.  We note that $ \frac{\ell^2}{4} e^{K} | \mathcal{K}| \geq 0$ for critical points, so that the timelike fibration in the metric (\ref{timelike_metric}) is stretched, and thus warped. This inequality is saturated only for the supersymmetric AdS$_3$ vacua, where no stretching occurs. Moreover, the Ricci scalar of the overall 3D spacetime, $R= 2 (4 T^2 + \mathcal{K} e^K)$, changes sign as one crosses this locus, an observation that justifies the billing ``de Sitter" in the internal region, but not de Sitter in the conventional sense, since the geometry is supersymmetric. Uplifting the warped critical points to 10D or 11D \cite{Cvetic:1999xp, Colgain:2014pha} one can show that CTCs appear for large values of $r$ \cite{Colgain:2015mta}. Finally, again suppressing $a_1$, we remark that there is an external locus, illustrated in FIG. 2, where critical points coalesce and only the supersymmetric AdS$_3$ vacuum exists
\be
a_2 = \frac{-1 + 2 a_3 -a_3^2 \pm \sqrt{a_3 -2 a_3^2 + a_3^4}}{2 (a_3-1)}. 
\ee

At (\ref{new_crit}) the Gaussian curvature may be written: 
\be
\label{K}
\mathcal{K} = 2 (a_1 a_2+ a_2 a_3 + a_3 a_1) - a_1^2 -a_2^2 -a_3^2 = \frac{2 a_1 a_2 a_3}{\ell}. 
\ee

For $\mathcal{K} < 0$, the critical points are easy to identify and correspond to supersymmetric G\"{o}del spacetimes \cite{Godel}, a healthy collection of which can be found in 3D \cite{Israel:2003cx,Banados:2007sq,Compere:2008cw,Levi:2009az}. To see this, we can recast the solution in the form \cite{Banados:2005da} 
\bea
\label{Godel}
\dd s^2_3 &=& - \left(\dd \tau + \frac{4 \Omega}{m^2} \sinh^2 \left( \frac{m \rho}{2} \right) \dd \varphi \right)^2\nn
&& \phantom{xxxxxxxx} + \dd \rho^2 + \frac{\sinh^2 (m \rho)}{m^2} \dd \varphi^2,
\eea
where in our notation, one has $\rho = \frac{2}{m} \tanh^{-1} (r)$, $\Omega = \frac{\ell}{4} | \mathcal{K}| e^{K}|_{\textrm{crit}}$, $m^2 = | \mathcal{K}| e^{K}|_{\textrm{crit}}$. 
Written in the above form (\ref{Godel}), the homogeneity and causal structure of the G\"{o}del solution holds in the range $ 0 \leq m^2 < 4 \Omega^2$ \cite{Reboucas:1982hn}, with the original G\"{o}del solution at $m^2 = 2 \Omega^2$ and AdS$_3$ at $m^2 = 4 \Omega^2$. We plot $4 \Omega^2 - m^2$ in FIG. 2, noting the yellow (zero valued) AdS$_3$ locus and steadily increasing contours outwards towards the boundaries. 
\begin{figure}[h]
\label{godel}
\centering
\includegraphics[width=0.3 \textwidth]{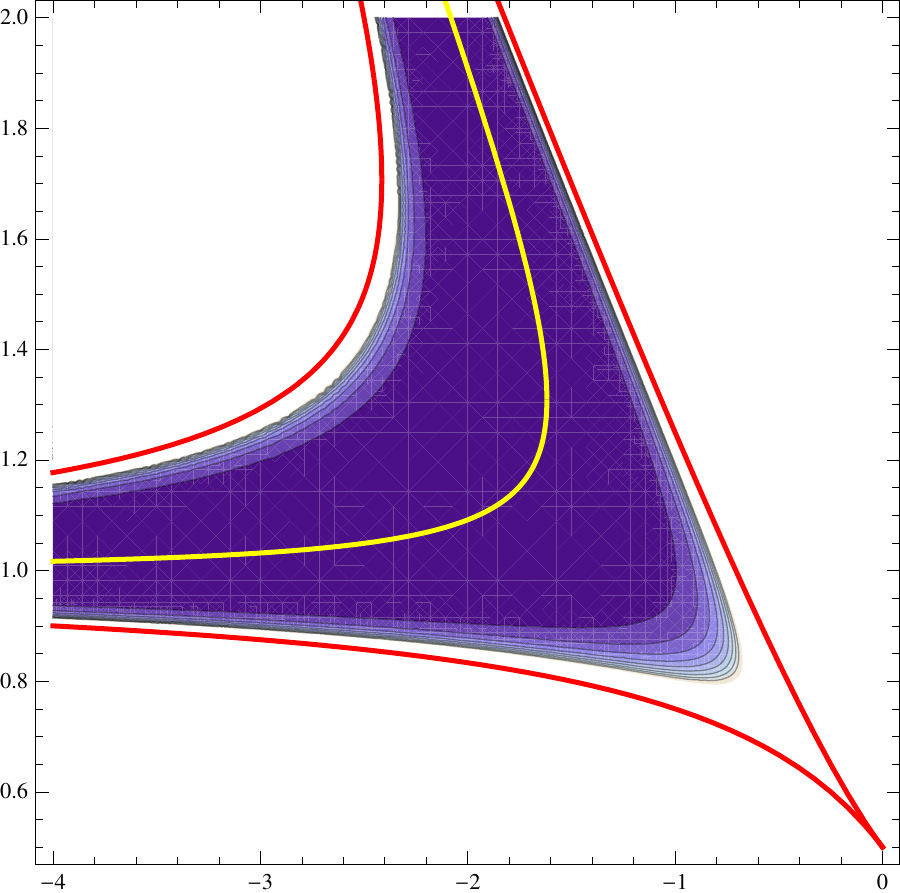}
\caption{A contour plot in the $(a_2, a_3)$ plane of stretching, $4 \Omega^2 -m^2$, in a sample $\mathcal{K} < 0$ region. Yellow curve corresponds to the AdS$_3$ locus. }
\end{figure}

For $\mathcal{N} = (0,2)$ SCFTs, the central charge is proportional to $T^{-1}$ \cite{Karndumri:2013iqa}, making it the natural candidate for a holographic $c$-function \cite{Freedman:1999gp, Berg:2001ty}. Indeed,  $T^{-1}_{\textrm{AdS}_3} > T^{-1}_{\textrm{G\"odel}}$, so for flows from AdS$_3$ to G\"{o}del, this observation suggests an analogue of Zamolodchikov's $c$-theorem \cite{Zamolodchikov:1986gt}. Recent work \cite{Compere:2014bia} on the asymptotic symmetries of warped AdS$_3$, including G\"{o}del, demonstrates that one can find two copies of the Virasoro symmetry, resulting in the expected central charge $c = \frac{3 \ell}{2 G}$ for a 2D CFT. Unfortunately, the analysis in \cite{Compere:2014bia} and earlier \cite{Compere:2007in} only holds for warped AdS$_3$ vacua where the cosmological constant does not change. In the language of 3D gauged supergravity, this means we are confined to warped AdS$_3$ vacua that co-exist with unwarped AdS$_3$ partners at constant value of the scalars in the potential. 

Here, our setting is more general, since the vacua exist at different values of the scalar fields, and these results do not apply. It is an open problem to repeat the analysis of Ref. \cite{Compere:2014bia} to see how the central charge depends on the scalar potential. Given the limitations of the literature, it is fitting to speculate that the inverse of the superpotential, as in the AdS$_3$ case, is the relevant quantity that encodes the central charge of the dual QFT. On this assumption \footnote{We adopt the same constant of proportionality as in the AdS$_3$ case.}, we can determine $c$ at G\"{o}del fixed-points in terms of twist parameters of $\mathcal{N} =4$ super-Yang-Mills: 
\be
c = 3 |\frak{g}-1| N^2 \prod_{i=1}^3 \frac{1}{a_i} (2 a_i^2 - \sum_{k=1}^3 a_k^2). 
\ee 
It will be interesting to repeat the analysis of Ref. \cite{Compere:2014bia} to determine the central charge for gauged supergravities. 

When $\mathcal{K}> 0$, little is known about these solutions, other than they exist as solutions to Topologically Massive Gravity \cite{Anninos:2009jt} and suffer from CTCs. Since they are topologically $\mathbb{R}~ \times$ H$^2$, it is possible that they can be analytically continued along the lines of \cite{Lin:2004nb} to give spacelike-warped AdS$_3$ with topology S$^1 \times$ AdS$_2$, on the proviso we change the sign of $T$.  We now show that this is not possible. To see this, we send $\dd s^2(\textrm{S}^2) = - \dd s^2(\textrm{AdS}_2)$  through redefining $r \rightarrow i \rho$, $\varphi \rightarrow i \tilde{\tau}$ and $\tau \rightarrow i \tilde{\varphi}$.  This would leave us with a signature problem, however this is overcome by $W_i \rightarrow \tilde{W}_i + i \pi$, an analytic continuation that allows (with $a_i \rightarrow - a_i$) flip the sign of $T$. Note, this leaves (\ref{scalarEOM}), (\ref{Liouville}) and (\ref{K}) unchanged. From the uplifted 5D perspective, this analytic continuation sends $\kappa \rightarrow -\kappa$, thus changing the genus $\frak{g}$ of the Riemann surface used in the 5D to 3D reduction. Unfortunately, the price one pays for this operation is that $G^i$ becomes complex, so the solution is not real. One can potentially overcome this by sending $T \rightarrow i T$, but then one sacrifices the consistent truncation, essentially by complexifying the theory. 

Before outlining the construction of numerical solutions in the next section, we end with a final remark that we have only discussed classical supergravity vacua and the $a_i$ should be quantised. To see this, we recall that the embedding in string theory is through a U(1)$^3$ fibration of S$^5$. For each U(1) isometry $\partial_{\varphi_i}$, the corresponding gauge field, $A^i$, must be a connection on a bona fide U(1)-fibration. This is equivalent to the condition that the periods of the first Chern class be integer valued, or 
\be
\frac{1}{2 \pi} \int_{\Sigma_{\frak{g}}} \dd A^i = 2 a_i (\frak{g} -1) \in \mathbb{Z}. 
\ee  

For $\frak{g} >1$ ($\kappa = -1$), where new critical points exist, this constraint poses little obstacle since we can ensure that the regions in FIG. 1 are populated by increasing the genus. 

\section{Supersymmetric Flows}
In this section we focus solely on parameters in the internal region of FIG. 1, where the timelike-warped de Sitter vacua exist, and construct a sample numerical solution to show flows from $\mathcal{N} = (0,2)$ fixed-points exist. In this region topology changes from H$^2$ to S$^2$, and $T$ changes sign making its $c$-function interpretation problematic. We note that linearising (\ref{scalarEOM}) about its AdS$_3$ values, there is an instability to flows in the direction of the timelike-warped dS$_3$ point. In contrast, flows to G\"{o}del are perturbatively stable.  
 
\begin{figure}[h]
\label{range}
\centering
\includegraphics[width=0.4\textwidth]{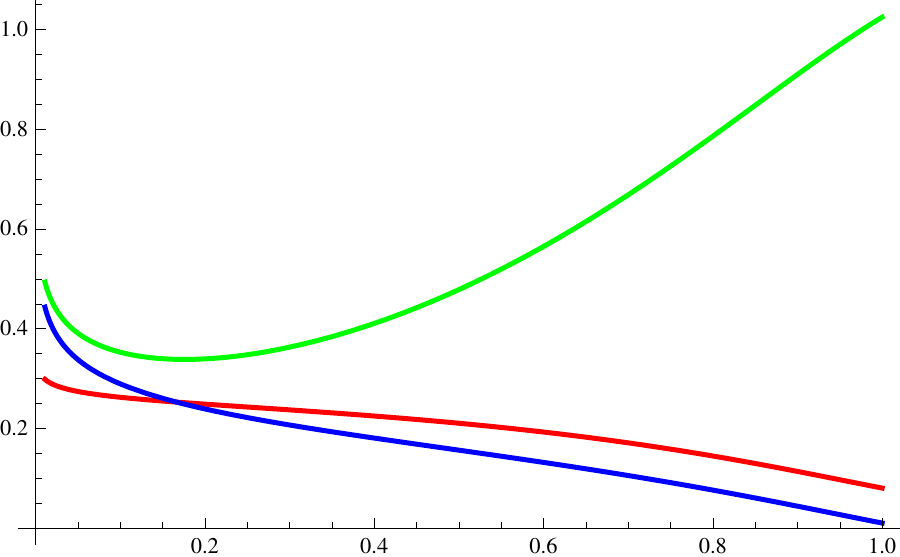}
\caption{Supersymmetric flow for $(a_1, a_2, a_3)=(\frac{3}{10},\frac{3}{10},\frac{2}{5})$, interpolating between AdS$_3$ values $W_{1} = W_{2}=\frac{3}{10}$ (Red), $W_{3} = \frac{9}{20}$ (Blue) at $r=0$ and warped AdS$_3$ values,  $W_1 =W_2 = \frac{2}{25}$, $W_3 = \frac{1}{100}$, at $r=1$. The Green curve corresponds to $D$. }
\end{figure}

Given a sample point in this region $(a_1, a_2, a_3) = (\frac{3}{10}, \frac{3}{10}, \frac{4}{10})$, we can use a shooting method, i.e. varying the initial conditions in the vicinity of $r=0$, so that the interpolating solution arrives at the second critical point at $r=1$. We have checked that the output of mathematica in FIG. 3 leads to an error of order $1 \times 10^{-7}$ when reinserted in the EOMs over the same range. Stiffness is encountered beyond $r=1$, but this is due to $T$ blowing up as $e^{W_i} \rightarrow 0$. We have linearised the scalar EOMs about the AdS$_3$ vacuum to extract the masses, $m^2_{\phi_{\pm}} \ell^2 = \frac{1}{2} (4\pm  3 \sqrt{3}) $, for scalars $\phi_{\pm} = \pm \frac{1}{\sqrt{3}} W_1 + \frac{1}{2} (1\mp \frac{1}{\sqrt{3}}) W_3$, and they are consistent with one relevant and one irrelevant operator. Tracing the fluctuation to the boundary of AdS$_3$ at $r=1$, it can be shown that the flows correspond to an irrelevant deformation of the SCFT \cite{Colgain:2015mta}. We anticipate a rich class of supersymmetric flows both between critical points. It remains to be seen how these solutions are related to black holes, so that the CFT interpretation can be elucidated. We hope to report on these in future work. 

We have benefitted from discussion with K. Balasubramanian, F. Bonetti, G. Comp\`ere, C. Herzog, K. Jensen, M. Ro\v{c}ek, M. M. Sheikh-Jabbari  \& H. Yavartanoo. We are grateful to J. Nian for earlier collaboration. E. \'O C acknowledges support from the Marie Curie grant ``T-dualities" and occasional hospitality from Port Jefferson Free Library.

\end{document}